\documentclass[]{mn2e}
\usepackage{psfig, epsf, epsfig}
\title[Formation of the Galactic GCs]{Formation  of the Galactic globular clusters with
He-rich stars in low-mass halos virialized at high redshift}
\author[K. Bekki, H. Yahagi, M. Nagashima,  and D. A. Forbes]
       {Kenji Bekki${}^1$\thanks{E-mail: bekki@bat.phys.unsw.edu.au},
        Hideki Yahagi${}^2$\thanks{E-mail: hyahagi@astron.s.u-tokyo.ac.jp},
        Masahiro Nagashima${}^3$\thanks{E-mail: masahiro@nagasaki-u.ac.jp},
        and Duncan A. Forbes${}^4$\thanks{E-mail: dforbes@astro.swin.edu.au} \\ 
        ${}^1$School of Physics, University of New South Wales,
              Sydney, NSW 2052,  Australia\\
        ${}^2$Department of Astronomy, University of Tokyo,
              7-3-1 Hongo, Bunkyo ward, Tokyo 113-0033, Japan\\
        ${}^2$Faculty of Education, Nagasaki University, Nagasaki,
              852-8521,  Japan \\
        ${}^4$Centre for Astrophysics \& Supercomputing,
Swinburne University of Technology,
Hawthorn, VIC 3122, Australia}

\begin{document}

\date{Accepted, Received 2005 May 13; in original form }

\pagerange{\pageref{firstpage}--\pageref{lastpage}} \pubyear{2005}

\maketitle

\label{firstpage}

\begin{abstract}

Recent observations have reported  that the Galactic globular
clusters (GCs) with unusually extended horizontal-branch (EHB)
morphologies show a significantly lower velocity dispersion 
compared with that of the entire Galactic GC system.
We consider that the observed distinctive kinematics
of GCs with EHB has  valuable information 
on the formation epochs of GCs
and accordingly  discuss this observational result
based on cosmological N-body
simulations with a model of GC formation.
We assume that  GCs in galaxies 
were initially formed in low-mass halos at high redshifts
and we investigate final kinematics of GCs
in their host
halos at $z=0$. 
We find that GCs formed in halos virialized 
at $z>10$ show  lower velocity dispersions on average
than those formed at $z>6$ for  halos
with GCs at $z=0$.
We thus suggest that the origin of the 
observed lower velocity dispersion
for  the Galactic GCs with EHBs is closely associated
with earlier formation epochs ($z>10$) of halos initially hosting
the GCs in the course of the Galaxy formation.
Considering that the origin of EHBs can be due to
the presence of helium-enhanced second-generation stars
in GCs, 
we discuss the longstanding second parameter problem
of GCs in the context of different degrees of chemical
pollution in GC-forming gas clouds within low-mass halos
virialized at different redshifts.

\end{abstract}

\begin{keywords}
globular clusters: general --
galaxies: star clusters --
galaxies:evolution -- 
galaxies:stellar content
\end{keywords}

\section{Introduction}

A growing number of observational studies  have recently reported that
some  Galactic GCs (e.g., $\omega$ Cen and NGC 2808)
show  a possible He abundance spread in their stellar populations
(e.g., Bedin et al. 2004; D'Antona et al. 2005; Lee et al. 2007, L07;
Piotto et al. 2007).
L07 furthermore have reported that
about 25\% of  Galactic GCs show very extended horizontal-branch (HB)
morphologies  that can be due to He-rich stars 
in these GCs.
One intriguing observation
of the Galactic GCs with extended HB morphologies (EHB) is that 
they show a significantly lower velocity dispersion ($93\pm13$ km s$^{-1}$) 
in comparison with the rest of the halo GC system 
($137\pm14$ km s$^{-1}$, L07).
Although this observation can shed  new light on the 
Galaxy formation through hierarchical merging 
of low-mass building blocks (L07),
the origin of the unique kinematics of the Galactic GCs
with possible He-rich stars remains unclear.

Although previous one-zone chemical evolution models have
discussed 
the observed possible star-to-star variation
in He abundance in Galactic GCs
(e.g.,  D'Antona et al. 2002; Bekki et al. 2007a),
they have not discussed the origin of the observed
kinematics of the GCs with EHB.
Recent numerical simulations of
GC formation during hierarchical galaxy formation
based on a $\Lambda$CDM model have greatly improved 
the predictability of 
dynamical and chemical properties of GC systems in
galaxies
(Kravtsov \&  Gnedin 2005;
Yahagi \& Bekki 2005; Bekki et al. 2007b). 
However they have not yet discussed the origin of
the observed lower velocity dispersion of  Galactic
GCs with EHB (hereafter these GCs are referred to as EHB GCs 
for convenience).

The purpose of this paper is to demonstrate,
for the first time, that the observed kinematics
of EHB GCs is consistent with a scenario in which
they were formed in low-mass halos virialized before
$z>10$.
We consider that  (i) the present-day metal-poor GCs of galaxies 
were formed initially in low-mass halos virialized at $z>6$ before 
the completion of reionization (Fan et al. 2003)
and (ii) GCs were then tidally stripped from their host halos
during hierarchical merging to  become
halo GCs in the final  galaxy.
The above (i) is consistent with recent hydrodynamical  simulations of GC
formation in low-mass halos by Bromm \&  Clarke (2002).
Based on high-resolution cosmological N-body simulations
with a model of GC formation,
we investigate the kinematics of GC subpopulations originating
from low-mass halos virialized at different redshifts ($z$).
We show that the  kinematical properties of  GC subpopulations initially
in low-mass halos virialized well before $z>10$ show 
significantly lower velocity dispersions in comparison
with  GCs formed at $z>6$. We discuss this result in
terms of the origin of EHB GCs in the Galaxy.

\section{The model}

Since the present model of GC formation in hierarchical galaxy formation
is essentially the same as those adopted in our previous
works (Yahagi \& Bekki 2005; Bekki et al. 2006; Bekki et al. 2007b),
we briefly explain the model here.
We simulate the large scale structure of GCs  
in a $\Lambda$CDM Universe with ${\Omega} =0.3$, 
$\Lambda=0.7$, $H_{0}=70$ km $\rm s^{-1}$ ${\rm Mpc}^{-1}$,
and ${\sigma}_{8}=0.9$ 
by using the Adaptive Mesh Refinement $N-$body code developed
by Yahagi (2005) and Yahagi et al. (2004), 
which is a vectorized and parallelized version
of the code described in Yahagi \& Yoshii (2001).
We use $512^3$ collisionless dark matter (DM) particles in a simulation
with the box size of $70h^{-1}$Mpc and the total mass 
of $4.08 \times 10^{16} {\rm M}_{\odot}$. 
We start simulations at $z=41$ and follow it till $z=0$
in order to investigate physical properties
of old GCs within virialized dark matter halos at $z=0$. 
We used the COSMICS (Cosmological Initial Conditions and
Microwave Anisotropy Codes), which is a package
of fortran programs for generating Gaussian random initial
conditions for nonlinear structure formation simulations
(Bertschinger 1995, 2001).

Our method of investigating GC properties is described as follows.
Firstly we select virialized dark matter subhalos at {\it each output} 
by using the friends-of-friends (FoF) algorithm (Davis et al. 1985)
with a fixed linking length of 0.2 times the mean DM particle separation.
The minimum particle number $N_{\rm min}$ for halos is set to be 10.
For each individual virialized subhalo,
the central particle is  labeled 
as a ``GC'' particle. This procedure for defining GC particles
is based on the assumption that energy dissipation via radiative cooling
allows baryon to fall into the deepest potential well of dark-matter halos
and finally to be converted into GCs.
We assume that  GC  formation is truncated 
for halos virialized
after $z=z_{\rm t}$ and that  $z_{\rm t}=6$ 
corresponds to the epoch of the completion
of reionization (e.g., Fan et al. 2003).
Secondly,  we follow the dynamical evolution of GC particles
till $z=0$ and thereby
derive locations $(x,y,z)$ 
and velocities $(v_{\rm x},v_{\rm y},v_{\rm z})$
of GCs at $z=0$.
We then identify virialized halos at $z=0$ with the FoF algorithm
and investigate whether each GC is within a halo.

We investigate velocity dispersions (${\sigma}_{z_{\rm v}<z}$)
of GC subpopulations
originating from halos  virialized
at redshifts $z_{\rm v}<z$ 
in  halos at $z=0$.
The most important key parameter in the present study
is the virialization redshift $z_{\rm v}$, 
and accordingly we investigate
the kinematics of GC subpopulations  
with $6 \le z_{\rm v} \le 15$ for $z_{\rm t}=6$
in the GC system (GCS)  in each of the simulated halos at $z=0$.
We particularly investigate the ratio  of 
${\sigma}_{z_{\rm v}<z}$ of GC subpopulations to 
the entire GCS 
in  each halo at $z=0$, denoted as 
${\sigma}_{\rm all}$.
These velocity  dispersion ratios  
are referred simply to as $R_{\rm s}$ from now on.
To avoid a large error bar 
of $R_{\rm s}$ (=${\sigma}_{z_{\rm v}<z}/{\sigma}_{\rm all}$
for  each individual halo at $z=0$)
resulting from the small GC number, 
we pick up haloes in which both the numbers of GCs originating 
from  halos with $z_{\rm v}<z$ and $z_{\rm t} < z \le z_{\rm v}$
exceed a threshold number of $n_{\rm th}$.  When
the total number of GCs in the $i$-th halo is $n_{\rm gc,i}$,
this criterion
corresponds to $n_{\rm gc,i} > 2 n_{\rm th}$. 
We mainly show the results for the models with $n_{\rm th}=3$
in which $n_{\rm gc, \it i}$ in each halo is larger 
than 6 ($=2n_{\rm th}$).
In order to discuss the dependences of the present results
on $n_{\rm th}$,
we also show the results of the models with  $n_{\rm th} = 24$.
Although an  error bar in $R_{\rm s}$ for each individual
halo becomes significantly small in the models with  $n_{\rm th} = 24$,
the total number of halos with  $n_{\rm th} \ge  24$
becomes very small.

L07 reported  that
the Galactic GCs with EHB 
has a velocity dispersion of $93\pm13$ km s$^{-1}$ ($={\sigma}_{\rm EHB GCs}$)
whereas the entire GCS of the Galaxy has
a dispersion of  $124\pm10$ km s$^{-1}$ ($={\sigma}_{\rm all}$).
We compare the observed ratio ($R_{\rm o}$) of
${\sigma}_{\rm EHB GCs}$ to ${\sigma}_{\rm all}$ 
(shown in  Table 1 of  L07)  for the Galactic GCS
with our simulated one in order to determine the best 
value of $z_{\rm v}$ for which the observed
$R_{\rm o}$ can be well reproduced.
Since the observed ${\sigma}_{\rm all}$ is derived for {\it all} GCs including
relatively young halo GCs (L07),
$z_{\rm t}=6$ is regarded as a reasonable
truncation epoch  in deriving ${\sigma}_{\rm all}$
for the present study.

The Galactic GCs around
the Galactic bulge (``bulge  GCs'' or ``disk GCs'')
could have been formed as a result of dissipative (and star-forming)
major merging possibly responsible for the bulge formation
at a lower redshift. The present model does not include
the formation of these  younger, metal-rich GCs, and accordingly
${\sigma}_{\rm all}$ is estimated only for GCs originating from halos
virialized before z=6. 
Thus it should be stressed that
the derived ${\sigma}_{\rm all}$ in the present model 
does not
literally mean the velocity dispersion of all GCs in a galaxy.
However we think that it is still reasonable to compare
the derived ${\sigma}_{\rm all}$ in the simulations with the observed
${\sigma}_{\rm all}$ (based on all GCs with known HB morphologies),
because the observed small number fractions
of bulge/disk  GCs in disk galaxies  (e.g.,
Zinn 1985 for the Galaxy)
mean that there could be  only a small difference
in velocity dispersions between a GCS including
bulge/disk GCs and the GCS not including them for a galaxy.

\begin{figure}
\psfig{file=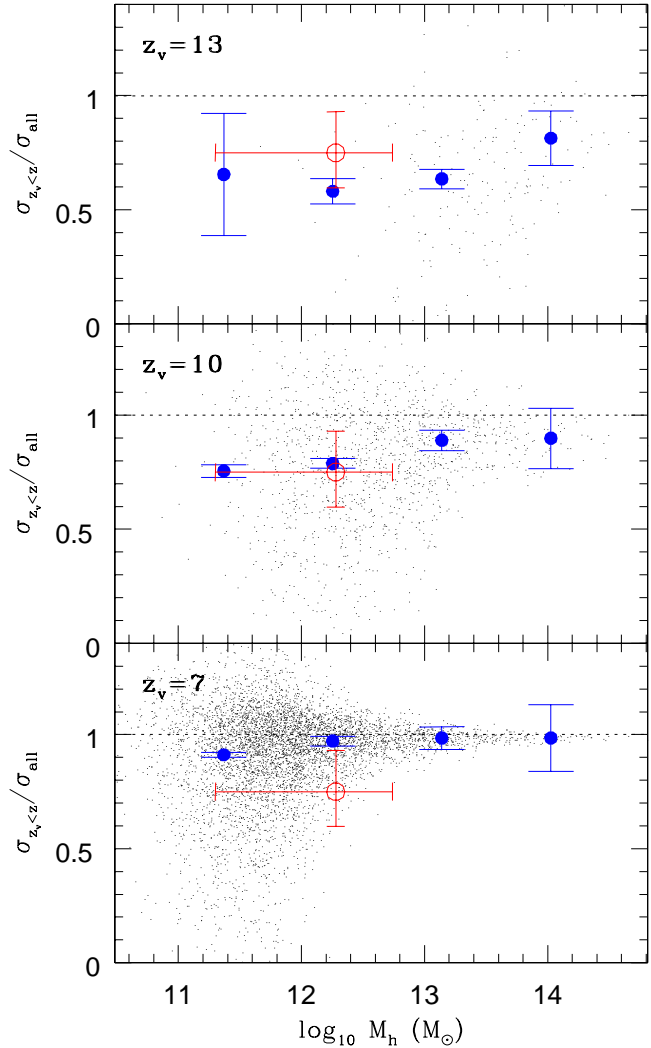,width=8.5cm} 
\caption{
Velocity dispersion ratios 
($R_{\rm s} = {\sigma}_{z_{\rm v} < z}/{\sigma}_{\rm all}$)
 as  a function of their host halo masses ($M_{\rm h}$)
for GC subpopulations  with
the virialization redshifts  $z_{\rm v}=13$ (top), 10 (middle), and 7 (bottom).
Here the ratios
are derived for the  $x$-components of velocity dispersion (${\sigma}_{\rm x}$)
in halos at $z=0$. 
Each small  dot represents a halo with a GCS with 
$n_{\rm gc, \it i}> 6$.  Blue,  filled circles represent the mean
values of the ratios in five  halo mass bins
and the error bars  are due to the number of halos ($N_{\rm h}$)
in each halo mass bin (i.e., $\propto 1/\sqrt{2(N_{\rm h}-1})$). 
For the lowest mass bin with no GCs for $z_{\rm v}=13$,
${\sigma}_{z_{\rm v} < z}/{\sigma}_{\rm all}$=0 is shown.
For comparison, the observed ratio
($R_{\rm o}$) of ${\sigma}_{\rm EHB GCs}/{\sigma}_{\rm all}$ (L07)
is shown by a red, open circle for the mass model of the Galaxy
(Wilkinson \& Evans 1999).
Note that GC subpopulations 
with $z_{\rm v}=10$
and $M_{\rm h} \approx 10^{12} {\rm M}_{\odot}$
show low ${\sigma}_{z_{\rm v} < z}/{\sigma}_{\rm all}$
which is  best agreement  to the observed one in L07.
}
\label{Figure. 1}
\end{figure}

\begin{figure}
\psfig{file=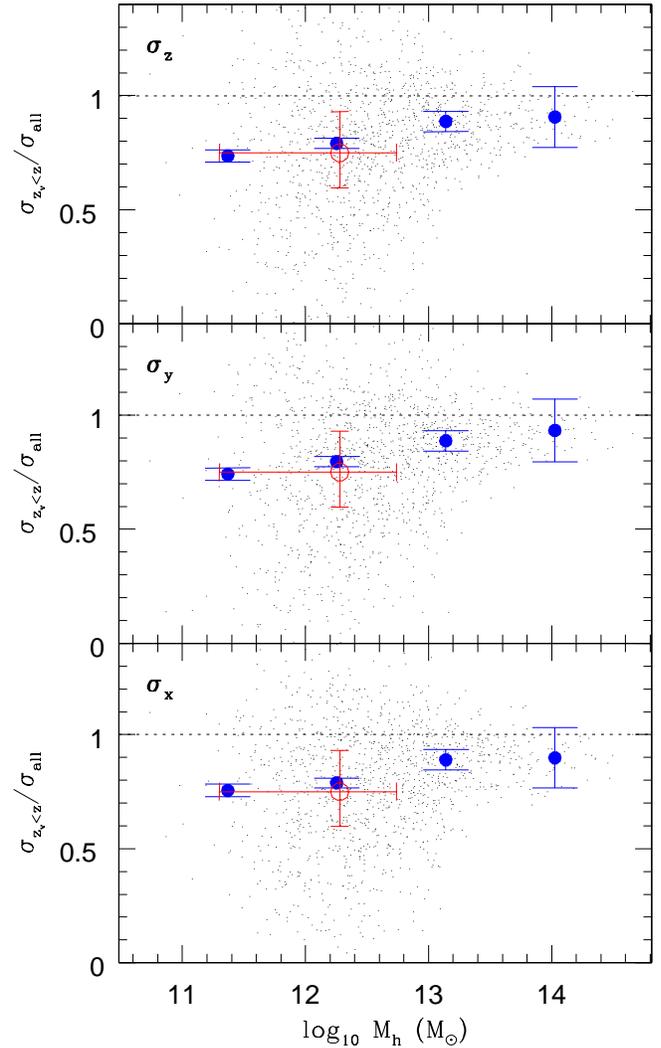,width=8.5cm} 
\caption{
The same as Figure 1 but for ${\sigma}_{\rm z}$
(top), ${\sigma}_{\rm y}$ (middle), and ${\sigma}_{\rm x}$ (bottom) 
of  GC subpopulations with $z_{\rm v}=10$.
}
\label{Figure. 2}
\end{figure} 

\begin{figure}
\psfig{file=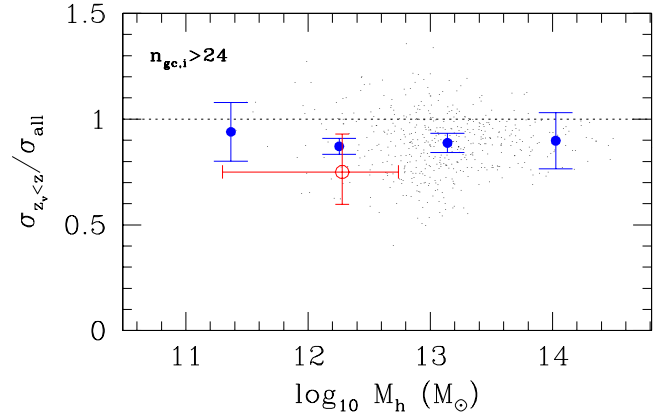,width=8.5cm} 
\caption{
The same as Figure 1 but for 
GC subpopulations  with $z_{\rm v}=10$ in 
the model  with $n_{\rm gc, \it i}> 24$.
}
\label{Figure. 3}
\end{figure} 

\section{Results}

Fig. 1 shows how $R_{\rm s}$ for  GC subpopulations 
with $z_{\rm v}$ = 7, 10, and 13 depend on
$M_{\rm h}$ in the model with $n_{\rm th}=3$
corresponding to $n_{\rm gc, \it i} > 6$.
Fig.1  demonstrates  that GC subpopulations with
higher $z_{\rm v}$ are more likely to show a lower 
$R_{\rm s}$  for a given $M_{\rm h}$,
though the dispersion in $R_{\rm s}$ for a given
$M_{\rm h}$ is larger  for low-mass halos.
This result suggests that $R_{\rm s}$ of a GC subpopulation
in a galaxy
can provide some information
about the redshifts of virialization for  low-mass halos
that initially hosted the subpopulation.
Fig.1 also  shows that more massive halos are more likely to show
a higher $R_{\rm s}$ value for a given $z_{\rm v}$,
which means that  differences in velocity dispersions between 
GC subpopulations originating from
high-$z$ halos with different redshifts of virialization
are less remarkable in more massive galaxies.

Furthermore, Fig. 1 shows that   GC subpopulations with
$z_{\rm v}=10$ in galaxy-scale halos
with $M_{\rm h} \approx 10^{12} {\rm M}_{\odot}$  
(corresponding roughly to the total mass of the Galaxy.
e.g., Wilkinson \& Evans 1999)
have  $R_{\rm s}$ very similar to  the observed velocity
dispersion ratio ($R_{\rm o}$):
The $R_{\rm s}$ of GC subpopulations with $z_{\rm v}=7$ 
are too high to be consistent with $R_{\rm o}$,
whereas those of GC subpopulations 
with $z_{\rm v}=13$ are  marginally
consistent with $R_{\rm o}$.
These results suggest  that EHB GCs
in the Galaxy  were initially within  low-mass
halos virialized at $z>10$ and later tidally stripped during 
the hierarchical merging of halos to finally become the halo
GCs in the Galaxy.
The origin of the  lower $R_{\rm s}$ in  GC subpopulations
with $z_{\rm v}=10$  may well be closely associated
with characteristic orbital properties (e.g, more eccentric
orbits and smaller pericenter distances with respect to
their halos' centers).

Fig. 2 shows that lower  $R_{\rm s}$ values for  GC subpopulations
with $z_{\rm v}=10$  can be clearly seen in the three
velocity  components 
(i.e., ${\sigma}_{\rm x}$,  ${\sigma}_{\rm y}$, and ${\sigma}_{\rm z}$)
in galaxy-scale halos with $M_{\rm h} \approx 10^{12} {\rm M}_{\odot}$,
which  confirms that lower $R_{\rm s}$ values
can be  one of the  kinematical properties characteristic of GC
subpopulations with  $z_{\rm v}=10$ in galaxy-scale halos.  
Fig.2 also shows that 
the trend of GC subpopulations 
in  more massive halos ($10^{13} {\rm M}_{\odot} \le M_{\rm h}$)
to have  higher  $R_{\rm s}$  values
can be clearly seen for the three velocity components.
This result implies that differences of
line-of-sight  velocity dispersions
in GC subpopulations 
(e.g., intra-group and intra-cluster GCs)
originating from halos virialized
at different redshifts are  observationally
difficult to detect for group-scale and cluster-scale
halos.

Fig. 3 shows $R_{\rm s}$ of halos in the model
with $n_{\rm th}=12$ 
corresponding $n_{\rm gc, \it i} > 24$ in which only halos with 
significantly larger numbers of GCs with $z_{\rm v}=10$
are selected.
Lower $R_{\rm s}$ values  in 
halos with $M_{\rm h} \approx 10^{12} {\rm M}_{\odot}$
can be clearly seen 
{\it even in  this strongly biased sample of halos at $z=0$},
though the scatter  in $R_{\rm s}$ become small 
thanks to a larger number of GCs used in deriving $R_{\rm s}$.
This result demonstrates that the derived lower $R_{\rm s}$ values
in galaxy-scale halos
are {\it not} due to the parameter $n_{\rm th}$ introduced
for minimizing scatters  in $R_{\rm s}$ for individual
halos in the present study.

\section{Discussion and conclusions}

\subsection{Origin of GCs with He-rich stars}

Recent observations have revealed that the 
(line-of-sight) velocity dispersions
of ``ultra-compact dwarf'' (UCD) galaxies 
are significantly smaller
than those of other  galaxy populations
in the Fornax and the Virgo clusters of galaxies
(Mieske et al. 2004; Jones et al. 2006; Gregg et al. 2007). 
Bekki (2007)  first demonstrated that
the  observed lower
velocity dispersion of the UCD population
in the Fornax cluster
is consistent with the UCDs having
significantly smaller pericenter distances  
(with respect to the center of the Fornax)
than other galaxy populations
in the cluster.
This result suggests that the simulated lower velocity dispersions
of GCs with $z_{\rm v}>10$ in halos
at $z=0$ are  due to significantly smaller pericenter distances 
of the GCs (with respect to the halos' centers)
in the present study.
It furthermore implies that EHB GCs 
in the Galaxy can have smaller pericenter distances
with respect to the Galactic center.

The present study has first shown that EHB GCs originate 
preferentially
from low-mass halos that were virialized at higher redshifts ($z>10$).
Given the fact that the origin of EHB can be closely associated
with the presence of He-rich second generation stars in GCs
(e.g., D'Antona et al. 2002; L07; but see Choi \& Yi 2007
for a diffferent formation mechanism of He-rich stars),
the above result implies that the formation of He-rich stars 
is much more likely to happen in GC-forming gas clouds 
of halos with possibly higher gas densities (due to higher $z_{\rm v}$).
Gas ejected from the first generation of stars of a  GC
can be efficiently trapped in the central regions 
of halos {\it during the GC formation}
owing to the deep gravitational potential of the halos
so that the gas can be used for the formation 
of the second generation with
a higher helium abundance for the GC (Bekki 2006):
there are two generations of stars with different helium
abundances in a GC.
Formation of He-rich second-generation stars 
therefore can proceed more efficiently in the central regions
of halos with
higher $z_{\rm v}$ and  thus deeper gravitational potentials
(for a given halo mass).
Therefore, formation of He-rich stars is highly likely to occur in 
GC-forming gas clouds located in the central regions
of halos with higher $z_{\rm v}$.

Although both observational and theoretical
studies  have discussed 
the origin of the second parameter(s)  that can control the morphologies
of HBs in GCs,
it is not yet clear what the second parameter(s) is (are)
in these studies (e.g., Catelan et al. 2001 and references therein).
Recently, Recio-Blanco et al. (2006, R06) have discovered that 
HB morphologies depend  on the total masses  of GCs 
in the sense that more massive GCs tend to have more extended HBs:
the masses of GCs can be one of the second parameters.
They have accordingly suggested that   higher fractions
of He-rich stars in more massive GCs
can cause EHB in the GCs and that  the origin of the 
higher fractions can result from more efficient ``self-pollution''
by stars formed earlier in more massive gas clouds.

If the number  fraction of He-rich stars in GCs is  one of the second 
parameters, as suggested by R06,
then the present numerical study suggests that
{\it the differences in HB morphologies in GCs are due partly
to the differences in  formation epochs 
(i.e., the redshifts of virialization) of low-mass  
halos initially hosting GCs.}
The present study furthermore suggests that if more massive
GCs with He-rich stars
are more likely to be formed in  halos
virialized at higher redshifts,
then the observational results
by R06 (i.e., the presence of EHB in massive  GCs) and by L07
(i.e., kinematic decoupling of EHB GCs)
can be self-consistently explained.
Owing to the lack of extensive theoretical studies on
GC formation processes in halos with different virialization redshifts, 
it is however unclear whether 
more massive GCs can be formed in halos virialized at higher redshifts.

It remains unclear whether AGB stars (e.g., D'Antona et al. 2002;
Karakas et al. 2006; Bekki et al. 2007) or  massive stars  (e.g., 
Charbonnel \& Prantzos 2006; Bekki \& Chiba 2007)
are responsible for the origin of He-rich stars in
GCs.  Miocchi (2007) has recently discovered an intriguing
correlation between the possible presence of intermediate-mass
black holes (IMBHs) and the presence of EHB in GCs.
Given the fact that recent numerical simulations 
(e.g., Portegies Zwart et al. 2004)
have demonstrated the formation of IMBHs through runaway collisions
of massive stars,
the above result  by Miocchi (2007) implies that 
massive stars rather than AGB stars 
could have played a major role in the formation of He-rich
stars within forming GCs.

\subsection{Chemo-kinematics of GC subpopulations and galaxy formation}

Although previous theoretical and
observational studies have focused mostly on
differences in kinematical properties between metal-poor and
metal-rich GCs in galaxies (e.g., Bekki et al. 2005;  Brodie \& Strader 2006),
they have not so far investigated 
the kinematical differences between GCs with different masses,
abundances of He and light elements (e.g., C, N, and O),
and $\alpha$-abundance ratios (e.g. [Mg/Fe]).
The discovery of the chemokinematical correlation
in the Galactic GCs by L07,
combined with the present numerical results
(i.e., lower velocity dispersions of GC subpopulations
originating from halos virialized at higher redshifts),
provide the following three implications.
Firstly,
there could be kinematical differences in GCs with He-rich
stars  compared to  those without in many  galaxies.
Recently, Kaviraj et al. (2007) have 
investigated the ultraviolet and optical  properties of 38 GCs in M87
and  suggested the possible
presence of He-rich stars  in the GCs.
It is thus an interesting observational question whether
the 38 GCs in M87 show a lower velocity dispersion in comparison
with the entire GC population in M87. 

Secondly,   velocity dispersions of
GC subpopulations with very high 
${\rm [\alpha/Fe] }$ (e.g., [Mg/Fe])
can be lower. This is mainly because
star formation time scales,
which determine ${\rm [\alpha/Fe] }$ of stellar populations 
(e.g., Matteucci \&  Fran\c cois 1992; see 
also Nagashima et al. 2005), 
is shorter in halos that were formed at  higher redshifts 
and thus have higher densities in their baryonic components. 
Thirdly, there can be significant kinematical differences
between GC subpopulations with lower masses and
those with higher masses in the sense that the latter subpopulations
should show lower velocity dispersions.
This is because the present simulations suggest that
more massive GCs can be formed at halos virialized at higher redshifts.

The observed chemokinematical correlations in the Galactic stellar
halo and the physical properties
of GCs have long been discussed in the context of
the time scale of the Galaxy formation 
(e.g., Eggen, Lynden-Bell, \& Sandage 1962)
and the accretion history of dwarfs with GCs
(Searle \& Zinn 1978).
The present study suggests that the observed kinematical correlations
of GCs with,  and without,  EHBs
in the Galaxy
can be considered in terms of different formation epochs 
of low-mass  halos initially hosting GCs.
We thus suggest that  HB morphologies of GCs in a galaxy  contain valuable
information both on the formation epochs of building blocks of the galaxy 
and on the hierarchical merging histories of that galaxy.
We also suggest that one of the second parameters
governing the HB morphology is the formation
(i.e., virialization) 
epochs of low-mass dark matter halos that initially
hosted  GCs at high redshifts.
In other words, the environments of GCs are a key `second parameter'
as they determine the extent to which GC-forming gas clouds are
chemically polluted by earlier generations of stars.

\section*{Acknowledgments}
%We are  grateful to the anonymous referee for valuable comments,
%which contribute to improve the present paper.
K.B. and  D.A.F. acknowledge the financial support of the Australian Research 
Council throughout the course of this work.
H.Y. acknowledges the support of the research fellowships of the Japan
Society for the Promotion of Science for Young Scientists (17-10511).
The numerical simulations reported here were carried out on 
Fujitsu-made vector parallel processors VPP5000
kindly made available by the  Center for Computational Astrophysics (CfCA)
at National Astronomical Observatory of Japan (NAOJ)
for our  research project why36b and uhy09a.

\end{document}